%% file: root.tex
\documentclass[letterpaper, 10 pt, conference]{ieeeconf}  

\IEEEoverridecommandlockouts                              

\input{packages}
\input{macros}

\usepackage{wrapfig}
\def\compileforpublish{1}

\begin{document}


\onecolumn

\title{\LARGE \bf
Designing an Automated Vehicle:  Strategies for Handling \\ Tasks of a 
Previously Required Accompanying Person*
}

\author{Tobias Schräder$^{1}$, Robert Graubohm$^{1}$, Nayel Fabian Salem$^{1}$, and Markus Maurer$^{1}$
	\thanks{\hspace{-1em}* This research was carried out as part of the UNICAR\emph{agil} project (FKZ 16EMO0285). We would like to thank the Federal Ministry of Education and Research (BMBF) for its financial support of the project and all members of the consortium for their contribution to this publication.
		\vspace{0.3em}}
\thanks{\hspace{-1em}$^{1}$Authors are with the Institute of Control Engineering at TU Braunschweig, Braunschweig, Germany.
        {\tt\small \{schraeder, graubohm, salem, maurer\}@ifr.ing.tu-bs.de}\vspace{0.3em}}
}

\maketitle
\thispagestyle{empty}
\pagestyle{empty}
%
\begin{abstract}%
\input{abstract}

\end{abstract}%

\IEEEpeerreviewmaketitle


\section{INTRODUCTION}
\input{introduction}
\section{NEED FOR RESEARCH}

\input{needforresearch}

\section{EXAMPLE SCENARIO}
\label{sec:scenario}
\input{example_scenario}

\section{IDENTIFIED STRATEGIES FOR FULFILLING SUPPORTING TASKS}
\label{sec:Design}

\input{ApproachforD}

\section{CASE STUDY: THE AUTONOMOUS FAMILY VEHICLE AUTOELF}
\input{casestudy}
\label{sec:casestudy}
\section{DISCUSSION}

\input{outlook}
\label{sec:discussion}

\section{CONCLUSION}
\input{conclusion}


\renewcommand*{\bibfont}{\footnotesize} 
\printbibliography

\end{document}

%% file: packages.tex
\usepackage{color}
\usepackage{tubscolors}
\usepackage{flushend}
\usepackage{tikz}
\usetikzlibrary{shapes,
				automata,
				arrows,
				arrows.meta,
				matrix,
				backgrounds,
				fit,
				patterns,
				decorations.markings, 
				svg.path, 
				shapes.multipart, 
				shapes.geometric, 
				external, 
				shadows, 
				patterns,
				positioning, 
				calc, 
				decorations, 
				scopes,
				fadings,
				bending,
				scopes}
\usepackage[simplified]{pgf-umlcd}
\usepackage{siunitx}
\usepackage[
			bibstyle=ieee,
			citestyle=numeric-comp,
			backend=biber,
			minnames=1,
			maxcitenames=2,
			maxbibnames=6,
			isbn=false,
			url=false,
			natbib=true,
			clearlang=true,
			urldate=comp
			]{biblatex} 
			\DefineBibliographyStrings{english}{urlseen = {accessed}}
\AtEveryBibitem{\clearlist{language}\clearfield{note}\clearfield{doi}}

\DeclareSourcemap{
	\maps[datatype=biber]{
		\map{
			\step[fieldsource=note, final]
			\step[fieldset=addendum, origfieldval, final]
			\step[fieldset=note, null]
		}
	}
}
\usepackage{xpatch}
\xpatchbibdriver{online}
{\printtext[parens]{\usebibmacro{date}}}
{\iffieldundef{year}
	{}
	{\printtext[parens]{\usebibmacro{date}}}}
{}
{\typeout{There was an error patching biblatex-ieee (specifically, ieee.bbx's @online driver)}}

\usepackage{ifthen}
\usepackage{lipsum}
\usepackage{amsmath}
\usepackage{amssymb} 	
\usepackage{bm} 		
\usepackage{booktabs} 	
\usepackage{graphicx}
\usepackage{microtype}
\usepackage{fmtcount}
\usepackage[hidelinks]{hyperref}



%% file: macros.tex
\makeatletter
\newcounter{IEEE@bibentries}
\renewcommand\IEEEtriggeratref[1]{%
	\renewbibmacro{finentry}{%
		\stepcounter{IEEE@bibentries}%
		\ifthenelse{\equal{\value{IEEE@bibentries}}{#1}}
		{\finentry\@IEEEtriggercmd}
		{\finentry}%
	}%
}
\makeatother

\newcommand\copyrighttext{%
	\footnotesize \parbox[t]{.11\textwidth}{\copyright{} \the\year~IEEE.} \parbox[t]{.89\textwidth}{Personal use of this material is permitted. Permission from IEEE must be obtained for all other uses, in any current or future media, including reprinting/republishing this material for advertising or promotional purposes, creating new collective works, for resale or redistribution to servers or lists, or reuse of any copyrighted component of this work in other works.}}
\newcommand\copyrightnotice{%
	\ifx \compileforpublish \undefined
	\else
	\begin{tikzpicture}[remember picture,overlay]
	\node[anchor=south,yshift=10.5pt] at (current page.south) {\parbox{\dimexpr\textwidth-\fboxsep-\fboxrule\relax}{\copyrighttext}};
	\end{tikzpicture}%
	\fi
}

%% file: abstract.tex
When using a conventional passenger car, several groups of people are reliant on the assistance of an accompanying person, for example when getting in and out of the car. For the independent use of an automatically driving vehicle by those groups, the absence of a previously required accompanying person needs to be compensated. During the design process of an autonomous family vehicle, we found that a low-barrier vehicle design can only partly contribute to the compensation for the absence of a required human companion. 

In this paper, we present four strategies we identified for handling the tasks of a previously required accompanying individual.  The presented top-down approach supports developers in  identifying unresolved problems, in finding, structuring, and selecting solutions as well as in uncovering upcoming problems at an early stage in the development of novel concepts for driverless vehicles.
As an example, we consider the hypothetical exit of persons in need of assistance. The application of the four strategies  in this example demonstrates the far-reaching impact of consistently considering users in need of support in the development of automated vehicles. 


%% file: introduction.tex
%

Driverless vehicles promise great opportunities for people who are unable to drive a conventional car on their own due to limited abilities. 
In this context, the potential of automatically driving vehicles has already been addressed by several companies in recent years. Waymo, by having a blind person ride in an automatically driving vehicle, illustrated opportunities associated with automated vehicles, for instance~\cite{iiiBlindManSets2016}. However, the use of an automatically driving vehicle in the real-world daily lives of people with or without limited abilities has not been widely field-tested.


When using a conventional passenger car, an accompanying person driving the car not only helps a person with support needs to overcome barriers that may be reduced by an adapted design. Due to limitations in their abilities, persons in need of support can be exposed to increased risks during the use of a car, for instance when getting in or out.  A human companion, for example a family member, is usually aware of this increased risk to the accompanied person and ensures their safety and well-being through his or her overall actions. 
The accompanying person performs the required tasks consciously or subconsciously even in unforeseeable situations and bears an objective or subjective responsibility (cf.~\cite{flemischDynamicBalanceHumans2012}).

A major requirement for an automated vehicle that is intended to be used by people who rely on assistance when using a car is that its passengers travel as safe as they have been with an accompanying person in order to maintain a positive risk balance. Therefore, it is necessary that the vehicle's passengers are not exposed to unreasonable risks even in unforeseen situations. A concept which meets these requirements is not just an automatically driving,  low-barrier vehicle that can be used self-reliantly, but a robot that can be entrusted to people in need of assistance. To achieve this  goal, it is necessary to compensate for the absence of a required accompanying person. Accordingly, supporting tasks which were previously carried out by an accompanying person cannot remain unfulfilled (cf.~\cite{schraderApproachRequirementAnalysis2019}). Furthermore,  responsibilities  previously performed by an accompanying person need to be redistributed.

Existing measures on both driverless and conventional vehicles -- such as low-barrier interior designs -- can already partially compensate for the absence of a person who previously performed supporting tasks. Depending on the type and scope of the supporting tasks, the existing solutions can be insufficient for certain use cases of an automated vehicle~\cite{schraderApproachRequirementAnalysis2019}. It can be assumed that requirements for existing design approaches and solutions for automated low-barrier vehicles are not explicitly formulated on the basis of the tasks performed by an accompanying person. At the same time, there is a lack of an approach that allows to systematically identify and compare the wide variety of possible design decisions that enable people with disabilities to use an automated vehicle independently.

In this paper, we present four   identified  strategies for dealing with the tasks of a nowadays needed accompanying person in the design of an automated vehicle. First, we present an overview of previous research on the use of automatically driving vehicles by  people with limited abilities  in Section~\ref{sec:needforresearch}. In Section~\ref{sec:scenario}, we describe possible supporting tasks of an accompanying person using a hypothetical example. Four strategies we identified to handle these tasks are explained in Section~\ref{sec:Design}. In Section~\ref{sec:casestudy}, we illustrate the application of these strategies in the design of an automated family vehicle. Subsequently, we discuss resulting key implications, challenges, and limitations for the development of new vehicle concepts.

%% file: needforresearch.tex
\label{sec:needforresearch}

Numerous requirements for vehicles  which are to be used by people who cannot use a conventional car independently are already known. Many of these requirements can be equally applied to current vehicles as well as to future driverless vehicles. At the same time, numerous individual solutions are already established -- especially for adults with motor and perceptual impairments -- that facilitate independent overcoming of barriers when using a vehicle. In this way, today's specifications for barrier-free public transportation can be met. Furthermore, there are also initial requirements for automated vehicles to be used by unaccompanied children. Many of the requirements we found in the literature were formulated from the perspective of parents.

Several authors have described the needs of people with physical impairments when using fully automated road vehicles. \textcite{tabattanonAccessibleDesignLowSpeed2019}, for instance, provide a review of documents describing the needs of elderly people and people with disabilities when using public transportation. One example for that is that the considered groups suffer from visual and motor impairments. These limitations imply special requirements for the geometric design of a vehicle's interior, for a vehicle's boarding system, or for the design of operating and information elements. \textcite{tabattanonAccessibleDesignLowSpeed2019} note that the lack of a bus driver who  assists passengers results in new requirements for automated vehicles. In this regard, the authors claim that there is a lack of standards for the design of accessible driverless vehicles. 

\textcite{gluckPuttingOlderAdults2020} describe the needs and desires of elderly, non-driving people for a shared driverless vehicle. The mentioned requirements include  abilities to store mobility aids, the option to request assistance in an emergency, a user-friendly design of the vehicle's controls, as well as an ergonomically designed ingress and egress. In addition, the desire for a door which is automatically actuated is described. Gluck \emph{et al.} see  benefits if the needs of the elderly  are considered in the design of shared driverless vehicles. 

In \cite{baylessDriverlessCarsAccessibility2019}, Bayless and Davidson describe the needs of people with disabilities regarding the use of automatically driving vehicles. They mention  problems that individuals with various disabilities -- whether temporary or permanent -- face when using current vehicles. They emphasize the importance of considering the needs of these groups in the design of automated vehicles. Therefore, \textcite{baylessDriverlessCarsAccessibility2019} suggest a Universal Design approach  and see great benefits in designing driverless vehicles accordingly. Bayless and Davidson also mention that new challenges in the operation of automated vehicles arise from the absence of a supporting person who is still present today.

The idea that a driverless vehicle could be used  by unaccompanied children has also been addressed by several authors. For example, \textcite{tremouletTransportingChildrenAutonomous2020a} examine factors that motivate parents of children between the ages of eight and 16 to let their children use a driverless vehicle on their own. In this context, the described requirements are  mostly functional and include two-way audio communication, an automatic locking system, an automatic notification of arrival at the destination, a verification that a restraint system is in place, and a system that checks the identity of a passenger. In addition, \textcite{leeAreParentsReady2020a}, \textcite{maFactorsAffectingTrust2020a}, \textcite{jingFactorsThatInfluence2021} and \textcite{koppelKeyFactorsAssociated2021}  describe similar requirements that parents express for a driverless vehicle. The mentioned authors examine factors influencing the acceptance of a fully automatically driving vehicle for transporting unaccompanied children.

A group of experts has  already formulated a list of requirements for driverless vehicles to be used independently by children~\cite{blueribbonpanelChildrenAutonomousVehicles2018}. However, the requirements  formulated predominantly address the special needs of children in terms of passive safety. At the same time, the report  identifies challenges arising from the absence of a supervisor and demands that this supervisor is to be replaced by technical measures.  

\textcite{ayoubOttoAutonomousSchool2020c} describe an autonomous school bus system that is designed to meet the needs of children and parents. They describe the design of entertainment and communication media and mention that a fully automated vehicle must ensure that its passengers can get off at a suitable place.

Children's requirements for a conventional passenger car driven by an adult companion are also well known. For example, \textcite{bubbAutomotiveErgonomics2021} describe special requirements of children for a road vehicle that go beyond passive vehicle safety aspects. For instance, it is mentioned that children have special requirements on the position of a door handle. Furthermore, children are prone to nausea, which results in special requirements for the vibration characteristics of a vehicle. A  geometrical design of a vehicle's window which enables children to have a potentially stabilizing eye-contact with the vehicle's environment can also be helpful to protect children from motion sickness.

Developers of the first driverless vehicles also considered the special needs of passengers with limited abilities. For example, some of the automatically driving vehicles introduced in recent years are equipped with a ramp and can be used by people in wheelchairs (e.g., \cite{SelfdrivingShuttleService2020, NationLargestFleet2021}). In this way, the vehicles can be used as public transportation and specifications such as the American  Disabilities Act ADA~\cite{ADAGovHomepage} are met. Another example is the vehicle Accessible Olli~\cite{ibmAccessibileTransportationIBM2018} -- an automatically driving shuttle that is intended to be used as a shared vehicle for shorter distances by people with limited abilities. A voice computer is used to enable communication between the vehicle and its passengers. However, despite these advances, this vehicle does not directly address the use case of accompanied rides in private cars and has a rather prototype character.

Overall, the requirements and solutions presented so far do not always meet the requirements that  arise from the omission of an accompanying person who is currently required when using a private car for accompanied trips~\cite{schraderApproachRequirementAnalysis2019}.  In the literature we reviewed, individual authors describe the absence of a previously present, supportive accompanying person as a challenge (e.g. \cite{tabattanonAccessibleDesignLowSpeed2019, blueribbonpanelChildrenAutonomousVehicles2018, baylessDriverlessCarsAccessibility2019}), however, it is usually not regarded as the central problem in the development of an automatically driving vehicle. One reason for this might be that  automatically driving vehicles assumed in the literature are often considered to be public transport vehicles rather than a replacement for today's private cars that are used to transport people who are unable to drive by themselves.  Although  supporting tasks of a bus or cab driver are also to be compensated for when designing an automated shuttle vehicle,  possible tasks  do not  extend as far as those of a person who transports his or her relatives in need of support in a private car.  Accordingly, the requirements formulated and solutions presented in the literature are not sufficient for the use case of an autonomous family vehicle, as described in~\cite{schraderApproachRequirementAnalysis2019}.

  For the complete automation of the pure driving task, the particular need to consider a vehicle as an overall system that takes over the complex driving task of a human is widely recognized. The diversity of  challenges caused by the fact that an automated vehicle is moving in an open world is a central and often discussed problem.  For  tasks performed by a person accompanying passengers in need of support when using a conventional car, such considerations are still unknown. 
  
  At the same time, the reallocation of complex tasks previously performed by humans alone between a human and a machine is not a new challenge. For example, basic principles and concepts for the realization of interaction between humans and robots are described by~\textcite{kraissBenutzergerechteAutomatisierungGrundlagen1998}. Frequently discussed application domains are e.g. aviation, production environments, or driver assistance systems. So far, however, the redesign and reallocation of the tasks of a supporting accompanying person when using a private car have not been considered in the literature we reviewed.  At the same time, no approach has yet been described that can be used to systematically translate the requirements resulting from perviously performed supporting tasks into design decisions.

%% file: example_scenario.tex
\begin{figure}
	\centering
	\includegraphics[width=0.6\columnwidth]{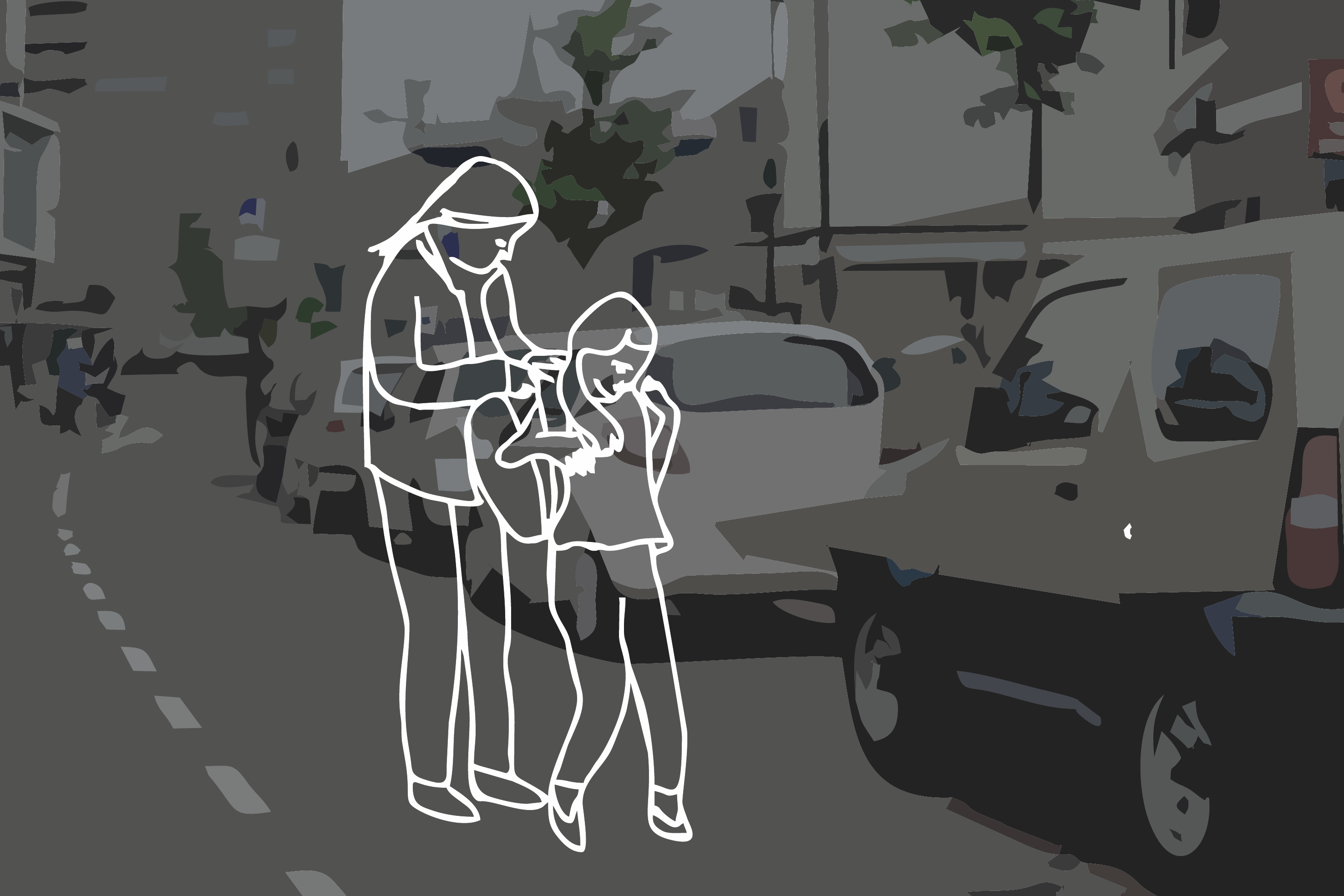}
	\caption{Illustration showing the support of a school child after the immediate exit from a car by an accompanying adult who ensures the child's safety}
	\label{fig:jung}
\end{figure}
\begin{figure}
	\centering
	\includegraphics[width=0.6\columnwidth]{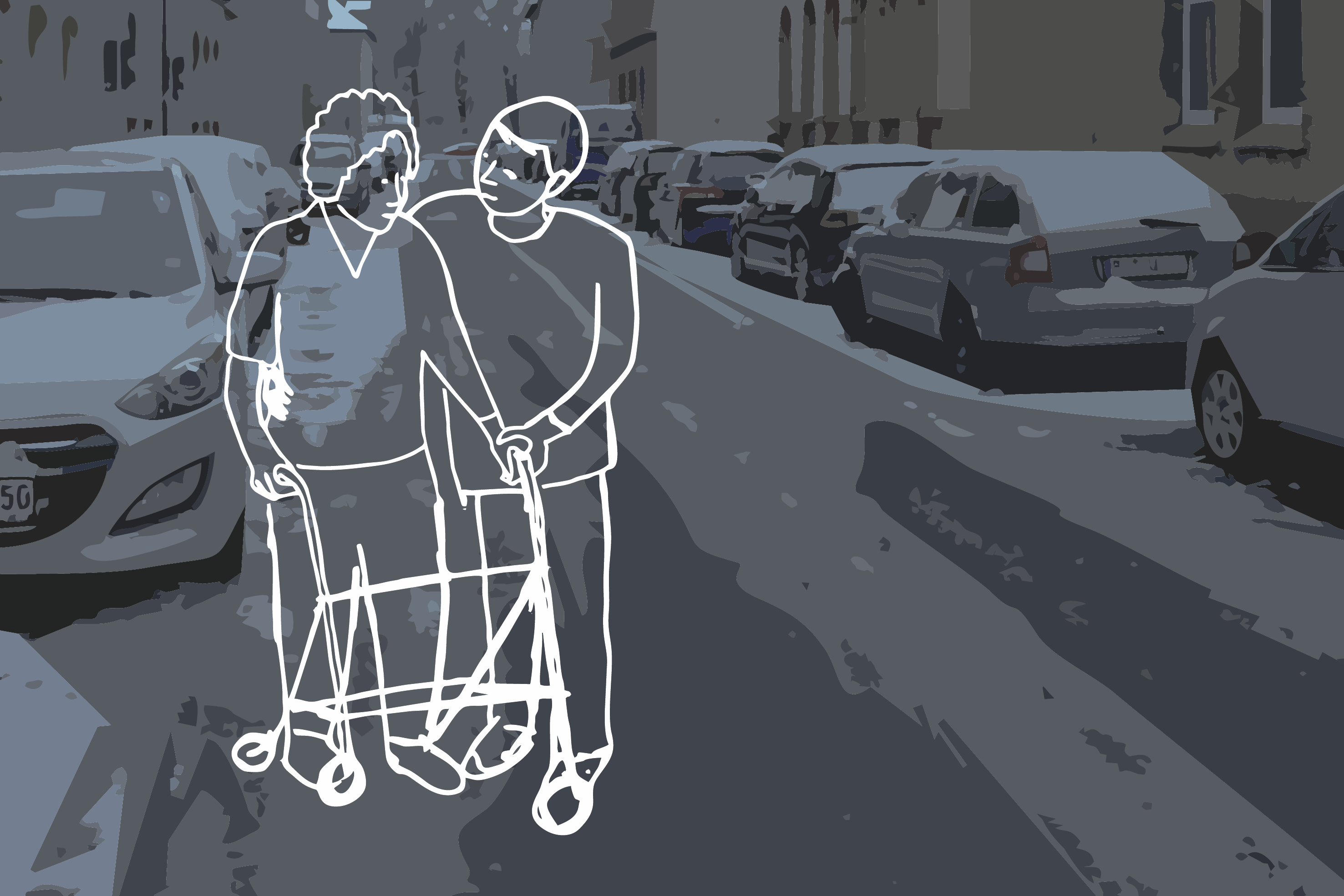}
	\caption{Illustration of accompanying an elderly locomotor-impaired person to the next sidewalk after leaving a passenger car on an icy road}
	\label{fig:alt}
\end{figure}

In order to illustrate the potential importance of an accompanying person, we introduce a sequence of a hypothetical accompanied trip in a conventional passenger car as an example. For this purpose, we choose the egress from a car.
As the accompanied person, we consider on the one hand a 12-year-old child who is let out of the car by a parent in front of the school (Fig.~\ref{fig:jung}). On the other hand, we consider an elderly person who is dependent on a mobility aid due to motor limitations (Fig.~\ref{fig:alt}). Both accompanied persons are not able to drive a conventional passenger car but  can walk shorter distances alone on the sidewalk under certain conditions. Both passenger's limitations include motor limitations that  make it difficult for them to get out of the car, to open the door, or to unload luggage. Moreover, both passengers have limited abilities to recognize hazards on the road. In case of the older person, the causes are limited perceptual and sensory-motor abilities. In the case of the child, the causes are  limited cognitive abilities~(cf. \cite{limbourgUberforderteKinderIm1998a}).

We assume that the car is parked in a longitudinal parking space near the destination by the accompanying person who is able to drive. When selecting the parking space, the accompanying person already takes into account whether the passengers he or she accompanies can walk from the place of exit to the actual destination. After the accompanying person has parked the car, he or she unloads the mobility aid or the school bag from the trunk.   Only after assessing the traffic situation, the accompanying person opens the door so that the accompanied person can get out. When overcoming the door sill, the accompanying person provides physical support to the individuals with motor impairments. While the older person needs time to position their mobility aid so that they can move around safely with it and the child puts on the school bag, the accompanying person continues to monitor the traffic situation. He or she ensures that longitudinal traffic does not pose a danger to neither the persons in need of support by guiding the accompanied person  and being  visible for other road users. Only when the accompanied person is safely on the sidewalk, the accompanying person leaves. He or she closes the trunk lid, gets in, and drives away.

For the initial scene of the hypothetical scenario introduced in this section, we assume that all of the accompanying person's tasks are necessary for a safe exit, since the abilities of the accompanied persons are not sufficient to reliably and independently perform the tasks involved during the sequence. Depending on external circumstances, further supporting tasks become necessary. For example, the ground could be icy, which means that the locomotor-impaired person
would need to reach the next cleared sidewalk.  The situation might be further complicated by a construction site blocking access to this sidewalk. 

Overall, the difficulty of the tasks involved in using the car, as well as an accompanied person's abilities, can be influenced by a variety of temporal factors. The type and intensity of support required is usually intuitively clear to the accompanying person. For the specification of a technical system, however, the variance of the tasks to be performed poses a particular challenge.

In \cite{flemischDynamicBalanceHumans2012}, Flemisch \emph{et al.} describe the kind of interaction between two human actors during the introduced scenario as a natural example of cooperative behavior, whose characteristics depend on the differences in the abilities, authority, control, and responsibility of the persons involved. 
In relation to the process described, this means that an accompanying person ensures that he or she has the necessary control during the egress procedure and is able to intervene. Accordingly, he or she has the required authority to act appropriately. Consequently, his or her role is coupled with a higher degree of responsibility for the accompanied persons. In  case of the older person, this responsibility can be just subjective, but in case of the child, it is also objective~(cf. \cite{flemischDynamicBalanceHumans2012}). If the persons supported in this example  are to use an automatically driving vehicle on their own, the accompanying person required today can no longer perform the tasks described in the current way and take the corresponding responsibility.


%% file: ApproachforD.tex

\begin{figure}
	\centering
	\includegraphics[width=0.55\columnwidth]{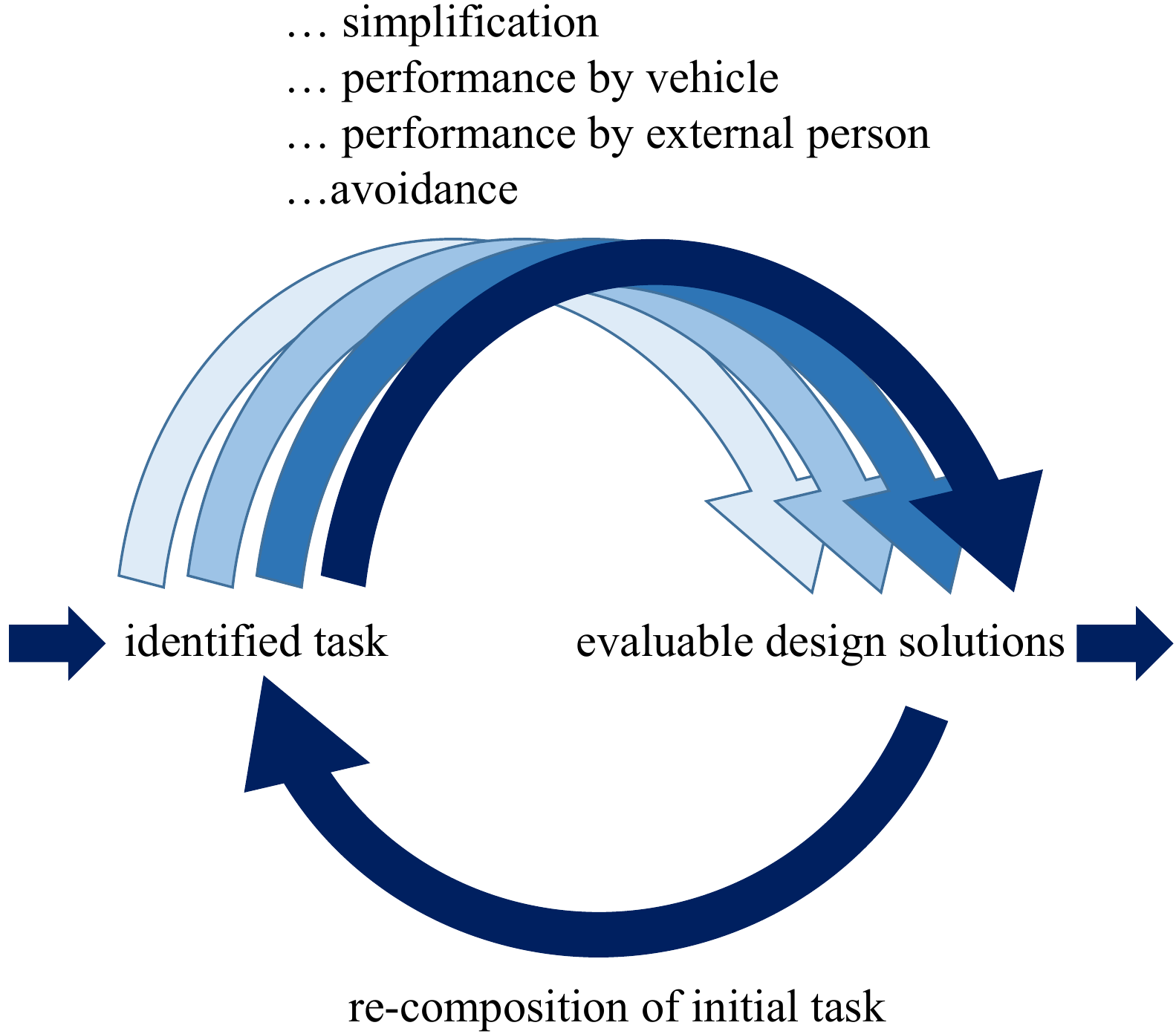}
	\caption{Basic procedure model: Application of the described strategies to an iteratively re-composable supporting task for the derivation of design solutions. }
	\label{fig:process}
\end{figure}

As mentioned at the beginning, the compensation for the absence of an accompanying person can be a prerequisite for the independent and safe use of an automated vehicle by some user groups. In~\cite{schraderApproachRequirementAnalysis2019}, we already presented an approach on how requirements for a driverless vehicle can be formulated based on an initial situation and how technical solutions can look like. The granularity of the individual tasks can vary and individual task an be decomposed into subtasks -- for example, in the course of an iterative procedure as depicted Fig.~\ref{fig:process}. In the three years since this initial publication in the beginning of the development phase of the automated family vehicle \emph{auto}ELF -- introduced in Section~\ref{sec:casestudy} -- four main strategies were identified for fulfilling the requirements derived from the tasks of an accompanying person. The concepts and contexts described in~\cite{flemischDynamicBalanceHumans2012} are  taken up in the following description of strategies for accomplishing tasks.  Decisions in favor of one of the strategies described below, i.e., the decision on how and by whom the tasks occurring during the operation of the vehicle are to be executed, is usually made by the vehicle developers. As sketched out in Fig.~\ref{fig:process}, an explicit application of the identified strategies emerges contrasting design solutions that can be evaluated and iterated by re-composing the underlying task in an early stage of a design process. 


\subsection{Simplification of Tasks}
\label{subesec:A}

By lowering the difficulty of a task and thus the demands on a person's abilities, passengers can  be enabled to perform a task themselves. Design principles like Universal Design -- which is suggested for example by \textcite{tabattanonAccessibleDesignLowSpeed2019} for the design of driverless vehicles -- or  design approaches for the compensation for users' limitations, for example described by \textcite{paetzoldApproachAdaptProduct2011}, can be assigned to this strategy.  Most solutions for driverless vehicles that were already presented, such as step-less entry, can be assigned to this strategy as well. As far as a passenger is able to perform a task independently, it is appropriate that this passenger is also given the authority to perform this task as well as the control and responsibility for fulfilling the task.

\subsection{Execution of Tasks by the Vehicle}
\label{subsec:B}
Not all supporting tasks that occur during vehicle use can be designed in  a way that they can be performed by a person riding with the vehicle.  For example, since children do not have  sufficient cognitive abilities to correctly assess a traffic situation \cite{limbourgUberforderteKinderIm1998a}, the required simplification of this task is not possible. Accordingly, the task of assessing the traffic situation must be taken over by another entity, such as the vehicle. In this case, the vehicle needs to have the ability to assess a traffic situation appropriately and it needs to be  aware of the necessity to fulfill this task. Furthermore, it needs to  behave in advance in  a way that it can fulfill this task -- if necessary. Depending on the differently distributed abilities, authority, control and responsibility are to be shifted to the vehicle respectively to its developers (cf.~\cite{flemischDynamicBalanceHumans2012}). 

\subsection{Execution of  Tasks by an External Person}
\label{subsec:C}
Tasks that can be performed neither by the vehicle nor by the passenger can still be performed by an external person.  Along with the ability to perform the task comes the assignment of control, authority, and responsibility to the regarded external person.  An external person can be enabled to perform tasks through technical measures. Accordingly, the nature of the transferable tasks depends  on the ability (including the provided resources) of an external person.  The execution of tasks by an external person can be part of the vehicle's regular operation. But in contrast to the other identified strategies, it is predestined as a fall back strategy. In the following Section~\ref{sec:casestudy}, we mention two different contrasting actors as examples of external persons who take over the diverse tasks of a former accompanying person.


\subsection{Avoidance of Tasks}
\label{subsec:D}
Another strategy to compensate for the absence of an accompanying person is to avoid tasks previously performed by him or her. This is possible, for example, by redesigning the procedures during the vehicle use, the vehicle's behaviors, the strategies used to achieve the mission objective, or a  restriction of the vehicle's operating conditions. In the case of the hypothetical accompaniment of a person on icy ground as illustrated in Fig.~\ref{fig:alt}, the task could be avoided by prohibiting the use of the vehicle in corresponding weather conditions. Therefore, vehicle use by persons requiring supportive tasks of an accompanying person that cannot be compensated for can also be avoided by limiting the operating conditions and formulating requirements for passengers.

%% file: Casestudy.tex
As part of the research project UNICAR\emph{agil}~\cite{woo18}, we aim to design a vehicle that takes over the task of accompanied trips in private car within a multigenerational family. Accordingly, this means that the vehicle  drives successfully children safely to school or takes elderly family members to a doctor's appointment. In this way, the vehicle, named \emph{auto}ELF, is intended to become an autonomous family helper. 

In the following section, we  illustrate the application of the described strategies to the prototype vehicle \emph{auto}ELF by giving a few examples. In this context, we refer to the supporting tasks of an accompanying person during the exit process described in Section \ref*{sec:scenario}. The solutions presented below were not selected without considering common compromises in the development of a vehicle, such as geometric constraints or weight targets -- which can be completely different in other vehicle projects. Moreover, it should be noted that this vehicle is a roadworthy prototype, but not all aspects relating to the use case are fully implemented yet. 

\begin{figure}
	\centering
		\vspace{0,2cm}
	\includegraphics[width=0.6\columnwidth]{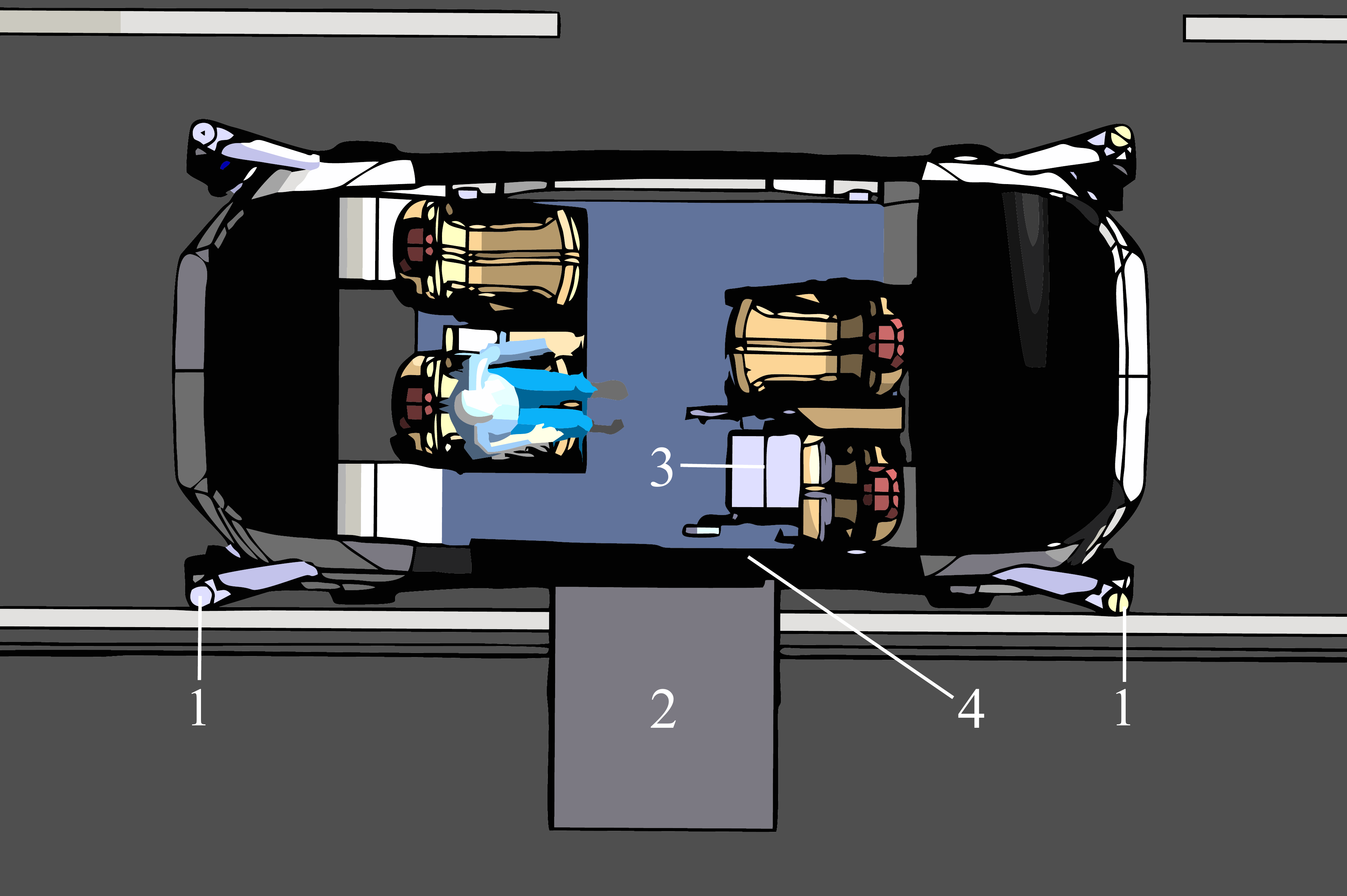}
	\caption{Top view of the \emph{auto}ELF vehicle~\cite{woo18}: 1: Sensor modules for detecting the vehicle's environment; 2: Lifting platform already extended in automated mode; 3: Mobility aid in partially automated holder device with seat folded up; 4: Automated door system (still closed) }
	\label{fig:Draufsicht}
\vspace{-0,5cm}
\end{figure}

An example for the simplification of a task (strategy \ref{subesec:A}) in the design of the autonomous family vehicle \emph{auto}ELF is the stowage of a mobility aid in the vehicle's interior. In the described initial scenario, the walking aid had to be unloaded from the trunk by an accompanying person. In the vehicle design we consider, the walking aid is stowed in the vehicle's interior as depicted in Fig.~\ref{fig:Draufsicht}. It is secured by  a semi-automatically activated rotary trap mechanism during the ride. The motor abilities required to unload the walking aid have thus been reduced.

The \emph{auto}ELF concept provides  strategy \ref{subsec:B} for the opening the doors. The doors of the vehicle open automatically~\cite{konigCONCEPTDEVELOPMENTPROCESS2021}. The vehicle's environment modeling is used to monitor longitudinal traffic in order to determine a suitable moment for opening the doors. The position of sensor modules, mounted on the outside (Fig.~\ref{fig:Draufsicht}) of the vehicle, enables to avoid unobservable areas.  Moreover, we assign the performance of all primary and secondary tasks of a person driving the vehicle according to the definition of~\textcite{bubbFutureApplicationsDHM2007} to strategy \ref{subsec:B} in order to compensate for the absence of an accompanying person. In~\cite{schraderCompensatingAbsenceRequired2021}, a draft of a functional system architecture for the fulfillment of tasks of an accompanying person by the automation of a vehicle was already presented. This strategy is not only bordered by the technical feasibility (e.g. in the area of machine perception and interpretation), but also by the lack of authority of the vehicle towards the passengers.

An example for the utilization of strategy \ref{subsec:C} is the transfer strategic decisions with a long-time horizon to an external person. In the concept of the autonomous family vehicle \emph{auto}ELF, we primary allocate planning tasks to  family members via a smartphone app. Thereby, we take advantage of a closed user group for the use case of a private vehicle  we consider. A major advantage of a passenger's relative is his or her familiarity with the passenger's specific needs and the trusted contact between both persons. Moreover, it can also be  a desire of a former accompanying person to fulfill planning and deciding tasks in the operation of an automated vehicle~\cite{tremouletTransportingChildrenAutonomous2020a}. Additionally, a control room that can even manually take over the driving task of the vehicle is foreseen as a fallback level for a variety of task. A major advantage of the control room in comparison to an external family member is its availability. 


An example for the avoidance of a task, described as strategy \ref{subsec:D}, is part of the entire egress procedure from the prototype family vehicle \emph{auto}ELF. The interior of the vehicle is designed in  a way that putting on a school bag does not need to take place outside the vehicle (cf. Fig.~\ref{fig:jung}) as in case of a contemporary passenger car. It is a prerequisite for  opening of the doors by the vehicle that there is no luggage in the storage compartments provided for this purpose in the vehicle's interior.  The task of securing the surroundings while the schoolbag is being put on is thus eliminated, since the school bag must be put in the protected vehicle's interior.  In general, the examples of task avoidance that we experienced during the prototype development led to inefficiencies -- in this case concerning the vehicle's geometry -- or to a reduction in the vehicle's usefulness. 

%% file: outlook.tex
The redesign and reallocation of the tasks of a present-day accompanying person pose new challenges in the development of automatically driving vehicles. A clear definition of the range of functions resulting from the tasks of an accompanying person is even more difficult as with the automation of the driving task. At the same time, the described strategies for fulfilling the tasks can have far-reaching effects on the diverse domains of vehicle design. The scheme described here opens a new perspective allowing to find solutions on a higher level of abstraction than described in previous work. The resulting advantages are basically the following advantages of abstractions, which are already described for the development of other complex systems(cf.~\cite{gorschek2006requirements}):

With regard to the development of an autonomous family vehicle described here, a concrete added value lies in the early clarification of the changed roles and associated tasks and responsibilities of individuals involved in the use of the vehicle under consideration. The clear allocation of the occurring tasks and the associated responsibilities is a prerequisite for a product that can be operated safely. Typical problems in the interaction between humans and machines -- e.g. described in~\cite{kraissBenutzergerechteAutomatisierungGrundlagen1998}  -- must be prevented. Particularly when considering passengers for whom someone else was previously responsible while using a conventional vehicle, the redistribution of tasks can be a decisive factor in the feasibility of a vehicle concept.

The application of the described scheme can help to find new solutions  and to question existing approaches. This can then lead, for example, to a comparison between functions, which were previously required solely on the basis of experience from a previous project, and a different definition of intended operational domains. 

Furthermore, the perspective described in this paper offers an opportunity to check specifications of novel vehicle concepts for their completeness: If a necessary task possibly performed during the use of a conventional vehicle by an accompanying person who is no longer present during operation of the intended automated vehicle  is not addressed by any of the described strategies, developers may have encountered a gap in their specifications.

The described considerations also indicate unanswered questions and upcoming challenges. We assume that some groups of people will continue to be unable to use an automated vehicle without human support.  Foreseeable misuse of the vehicle  by transporting ineligible passengers must also be addressed by defining responsibilities between the vehicle's developers  and the user who allows his or her relatives in need of support to use the vehicle by themselves. An option to minimize  the risk of harm to passengers from unintentional misuse of the vehicle is to define and clearly communicate the limits of the vehicle's abilities. Thereby, it must also be ensured that tasks which the vehicle cannot adequately perform are not assigned to the vehicle developer's area of responsibility in the event of an accident.

 A complete list of requirements arising from the wide range of tasks performed by an accompanying person cannot be expected. At the same time, the number of possible test cases for the release of a vehicle in the sense described here is not predictable, yet.  Unpredictable behavior of the vehicle's users, who entrust or are entrusted to it, adds another dimension of uncertainty. If previous approaches for the validation of automated vehicles and the creation of test scenarios are sufficient for the required functional scope of a vehicle that assumes the tasks of an assisting person needs to be clarified. In this context, evaluating the difficulty of a task is also challenging. The extent to which known approaches (e.g. NASA-TLX~\cite{hartNasaTaskLoadIndex2006}) are sufficient for this purpose is unclear. Thereby, a single test vehicle, such as the prototype we developed as part of the  UNICAR\emph{agil} project, is not sufficient for a complete validation. At this point, it should be noted that our considerations are limited to the early development stage of a novel vehicle. The concept proposed in this paper is not intended to replace usability testing or other real-world testing. At the same time, the completion and evaluation of the described prototype is pending.

With the previously listed uncertainties in mind, we recommend to ask at an early stage in the development of new automated vehicle concepts which tasks a former accompanying person performed and what it means to simplify these tasks, to have them performed by the vehicle, to have them performed by an external person or to avoid them. Furthermore, we recommend to see it as an indicator for a gap in the specifications of a vehicle concept, if a formerly required tasks of an accompanying person is not addressed by any of the mentioned strategies. 

%% file: conclusion.tex
To compensate for the absence of  accompanying persons, supportive tasks that they have been performing must be redesigned, reallocated, or avoided. At the same time, their previous responsibilities must  be redistributed. In this paper, we  present four strategies for handling the tasks of a required accompanying person when using a vehicle. A consideration of these strategies reveals a wide variety of design choices that enable independent vehicle use for individuals who otherwise rely on human assistance. A low-barrier, inclusive vehicle design represents only one part of the options. The individual strategies that vehicle designers can choose have various advantages and disadvantages that depend on the intended use case of the vehicle under development. The type of strategy chosen to accomplish an accompanying person's task can have a far-reaching impact on the design and subsequent usage of an automated vehicle. Explicitly considering different ways of dealing with required supportive tasks in an early stage of designing a new type of vehicle can add value to the resulting overall system.  The consideration of selected tasks of an accompanying person and the attempt to avoid or fulfill these tasks by the design of an experimental vehicle indicate the complexity and extent of upcoming challenges in this context.
